\documentclass[sigconf]{acmart}

\usepackage{booktabs} 
\usepackage{listings}
\usepackage{graphicx}  
\usepackage{url}        
\usepackage{amsmath}   
\usepackage{caption}
\usepackage{color}
\usepackage{comment}
\usepackage{physics}
\usepackage{makecell}
\usepackage{array}
\usepackage{multirow}
\usepackage{diagbox}
\usepackage{dirtytalk}

\copyrightyear{2019}
\acmYear{2019}
\setcopyright{acmlicensed}
\acmConference[FIRE '19]{Forum for Information Retrieval Evaluation}{December
12--15, 2019}{Kolkata, India}
\acmBooktitle{Forum for Information Retrieval Evaluation (FIRE '19), December 12--15,
2019, Kolkata, India}
\acmPrice{15.00}
\usepackage{diagbox}
\acmDOI{10.1145/3368567.3368583}
\acmISBN{978-1-4503-7750-8/19/12}

\begin{document}
\title[Information Foraging for Enhancing Implicit Feedback in Content-based Image Recommendation]{Information Foraging for Enhancing Implicit Feedback in Content-based Image Recommendation}

\author{Amit Kumar Jaiswal}\thanks{This work is supported by the Quantum Access and Retrieval Theory~(QUARTZ) project, which has received funding from the European Union's Horizon 2020 research and innovation programme under the Marie Sklodowska-Curie grant agreement No. 721321.}
\affiliation{
  \institution{University of Bedfordshire}
}
\email{amitkumar.jaiswal@beds.ac.uk}

\author{Haiming Liu}

\affiliation{
  \institution{University of Bedfordshire}
}
\email{haiming.liu@beds.ac.uk}

\author{Ingo Frommholz}
\affiliation{
  \institution{University of Bedfordshire}
}
\email{ifrommholz@acm.org}


\renewcommand{\shortauthors}{Jaiswal et al.}

\begin{abstract}
User implicit feedback plays an important role in recommender systems. However, finding implicit features is a tedious task. This paper aims to identify users' preferences through implicit behavioural signals for image recommendation based on the Information Scent Model of Information Foraging Theory. In the first part, we hypothesise that the users' perception is improved with visual cues in the images as behavioural signals that provide users' information scent during information seeking. We designed a content-based image recommendation system to explore which image attributes (i.e., visual cues or bookmarks) help users find their desired image. We found that users prefer recommendations predicated by visual cues and therefore consider the visual cues as good information scent for their information seeking. In the second part, we investigated if visual cues in the images together with the images itself can be better perceived by the users than each of them on its own. We evaluated the information scent artifacts in image recommendation on the Pinterest image collection and the WikiArt dataset. We find our proposed image recommendation system supports the implicit signals through Information Foraging explanation of the information scent model.


\end{abstract}

%
%

\ccsdesc[500]{Information systems}
\ccsdesc[500]{Information systems~Personalization}
\ccsdesc[500]{Information systems~Recommender systems}
\ccsdesc[300]{Human-centered computing~Web-based interaction}

\keywords{Information Foraging Theory, Image Search, Recommendation System}

\maketitle

\section{Introduction}
Personalised recommendation systems assist users in dealing with information overload by informing them about relevant items. When it comes to image recommendation, earlier studies~\cite{messina2019content} have relied on metadata and some work has leveraged visual features extracted with deep learning driven neural networks to recommend art. However, the problem of information overload~\cite{sundar2007news,kim2016role} in an image search (or recommendation) scenario has been studied by adjusting or enhancing the search interface~\cite{wilson2008improving} rather than by leveraging implicit feedback improving recommender systems by learning user behaviour signals. In this work, we contribute to the area of behavioural frameworks applied in content-based image recommender systems by exploring Information Foraging Theory (IFT)~\cite{pirolli1999information}, which is a theory originating from cognitive psychology that helps users to navigate and forecast their behavior through the information environment when searching relevant information. Users usually seek to put less effort into their information seeking process and expect to maximize their information gain. Information Scent, one of the constructs of Information Foraging Theory, was used to manipulate search engine result pages~(SERPs) deliberately by inspecting changes in web search behavior. Earlier studies~\cite{wu2014online} discovered that detail observation of documents and clicking deeper in search results leads to stronger scent. 

In this work, we investigate if Information Foraging Theory can be applied to image recommendation systems by encompassing implicit feedback signals. We evaluate the strength of information scent using implicit signals (i.e., visual cues), which corresponds to the search task relevance. Our evaluation methodology utilises 
the image collection from a popular visual discovery engine - Pinterest.com. We explore how visual cues affect user preferences by estimating their information scent. 
\vspace*{-2mm}
\section{Related Work}
This section covers studies on Information Foraging Theory and content-based image recommender systems.
\vspace*{-2mm}

\subsection{Information Foraging Theory} Information Foraging Theory~(IFT), developed by Pirolli and Card~\cite{pirolli1999information}, stems from the ecological science concept of optimal foraging theory as employed to how humans seek for information. IFT provides stipulated constructs borrowed from optimal foraging theory which includes how predators conform to humans who seek for information (or prey). They delineate these searches for instance in user interface sections, called patches. From a foraging perspective in image search, the searcher is the predator, the information patch is any segment or a region within an image in an artifact of the environment, the piece of information a user looks for is the prey, and the consumed~(or gained) information is the information diet. 
According to IFT, there are three different models which describe various aspects of user behaviour (foraging) for information:
    \emph{Information Patch} models deal with time allocation and information filtering activities.
    \emph{Information Scent} models help people make use of perceptual cues, such as Web links spanning small snippets of graphics and text, consecutively to make their navigation decisions in selecting a specific link. The purpose of such cues is to characterise the contents that will be envisaged by trailing the links. 
    \emph{Information Diet} models deal with decisions about the combined set of information that has some perceived value to a searcher, who then pursuits with that set of information and neglects the remaining part. 
Information scent manifests the proximal cues and distal information sources they lead to (i.e., their perceived relevance). This theory infers that users must seed their navigation decisions on evaluation of information scent cues allied with individual choices. The user will evaluate the link likely to lead his/her information goal if the link cue is perceived to have high information scent. 

IFT has been applied on text data to model users' information need and their actions using information scent~\cite{chi2001using}. However, it has been previously found that information scent can analyse and predict usability of a website by determining a website's scent~\cite{chi2000scent}. Liu et~\textit{al}.~\cite{liu2010applying} demonstrated an IFT-inspired user classification model for content-based image retrieval systems to understand the users' interaction by functioning the model on several interaction features collected from the screen capture of different user task types. They evaluated their classification model by performing qualitative data analysis and found that the six characteristics in the model are consistent with those interaction features which built a preliminary practice to study user interaction/behavior via Information Foraging Theory. Recent work~\cite{schnabel2019shaping} studies the influence of feedback data in a movie recommendation system by altering the user interface using information scent and information access cost and found that the primary task of selecting a movie to watch improves the implicit feedback data. 
\vspace*{-3mm}
\subsection{Implicit Feedback in Recommender Systems}
Several recent works in the recommendation context focused on user-generated contents such as user tags and forum comments, which were leveraged in many ways, e.g., to improve item and user profiles, or to explain recommendations to the user~\cite{jaschke2007tag,vig2009tagsplanations}. However, researchers focusing on user-provided tags tried to extract different types of information which include user opinions, semantics and sentiments, that can be processed in the recommendation process. Also, tag-based recommender systems rely on user generated content not only to exploit this information to enhance the quality of recommendation, but also as a measure to explain the recommendations to users. People still find it challenging to effectively exploit multimedia content. Recent work~\cite{berget2019textual} found that textual search interfaces are not sufficient enough to follow users' cognitive skills. However, multimedia recommendation heavily rests on explicit and implicit user feedback. To this end, the crucial part in recommending multimedia content is about feature extraction, and how they should be integrated with user preference data and user generated content to confer the most relevant recommendations~\cite{hernandez2019comparative}. 

Implicit user feedback establishes observable behaviours presented by a user. Past work~\cite{oard2001modeling} details useful behaviour, categorised in four classes which includes examine, retain, reference and annotate. They introduced a classification framework for recommender systems to compare explicit and implicit user feedback with a set of specific properties, which they utilise to model user preferences in items. However, our work focuses on the \say{examine} property of observable behaviours.

\vspace*{-2mm}
\section{Content-based Image Recommendation Engine (CBIR)}
\label{sec:cbir}
\begin{figure}[h!]
  \includegraphics[width=\linewidth]{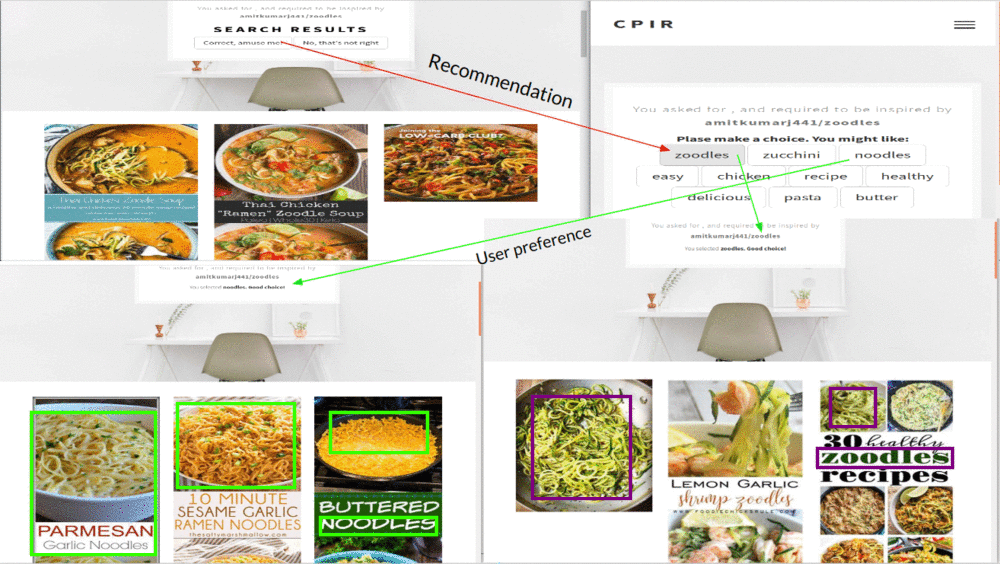}
  \caption{\textbf{Content-based Image Recommendation Interface}}
  \label{fig:pgui}
\end{figure}

We develop a content-based image recommendation engine for image search based on the User-Image-Cue model~\cite{jaiswal2019effects}. The User-Image-Cue model is framed with an interlinked graph consisting of users, images and re-ranked cues set. The search interface with recommendations is reported in Figure~\ref{fig:pgui}.
The advantage of using content-based recommendation over collaborative filtering is that it does not have the cold start problem~\cite{lika2014facing}, where a new item (or user) is introduced without prior history as well as ample amount of data, whereas the latter leverages the users' correlation to make a recommendation. The content-based recommendation engine directs the image objects in the representation space, which allows for recommending similar items.

We design our recommendation system by adding a Pinterest board widget which lets the user input his/her board name\footnote{A board allows people to place all of their visual cues around diverse interests, ideas and plans} as a query from Pinterest based on keywords and it mines the entire image collection in real-time from the specific board; users, while seeing the search result, will be given several preferences based on their current search result to choose from, and if selected, the recommendation system again returns similar items. Each and every image from the collection includes a cue associated with it. Recent work~\cite{jaiswal2019effects} found that such a system exhibits foraging\footnote{Foraging phenomena in this work refers to Information Foraging Theory} effects in personalising recommendation.

\vspace*{-2mm}
\section{Implicit Feedback Signals}\label{sec:sec_4}
\subsection{Cues as Artifacts of Information Scent}
In Information Foraging Theory, the flux of information in an information environment perceives user plans for seeking, gathering and consumption based on their utility~\cite{pirolli1999information}. The view of animals following scents to forage is analogous to users following various kinds of cues, assessing information contents and navigating across information spaces depending on their information scent. Cues cater as proximal cues in order to emit information scent. It is usually represented by web elements which can be recognised as a visible representation of users' mental beliefs. In other foraging tasks, by enabling these mental beliefs, it confer the incomplete realization of the user access path of information content or utility (cost and benefits)~\cite{pirolli1999information}.

We consider images as exemplary web elements (more or less an information patch) that can be attended via cues. Viewing some web sources enables users' cognitive beliefs. A user may activate such key beliefs via cues, after perceiving ideas and plans for seeking.

\subsection{Using Cues for Information Scents}
Cues not only ratiocinate the internal mental representations but also confer an external information scent to reconstruct representations in successive information processing tasks. Cues are often commonly generated deliberately and prospectively to minimise the information access cost of successive operations. Eventually, the quality of a cue depends on how efficiently and effectively it helps later information foraging, recommendation, and information retrieval.

We investigate the quality of cues for two different kinds of usage: recommendation system and information retrieval~\cite{fu2008microstructures}. For recommendation systems, cues provide user preferences for an image, and generate an information scent to spot related interesting images~\cite{loumakis2011image}. These cues provide incomplete information to decide on navigation paths in real-time to pursue information. For information retrieval~\cite{chi2000scent}, cues provide information scent to spot the document (e.g. keywords to search) and renovate activations that have decayed. 

Users can gain different information scent from identical cues. The same user can also update his/her cues with reconstructed knowledge gained throughout the information seeking tasks. To gain some preliminary insights, we ran a pilot user study using our CBIR system that we presented in Section~\ref{sec:cbir} with three participants on Pinterest and WikiArt~\cite{saleh2015large} collection. Our proposed recommendation system (Figure~\ref{fig:pgui}) incorporates visual and textual cues so that the user is able to pick an interesting image from the recommended list of items~(or via a list of recommended user preferences). The selection of images by the user is supported via cues trails which are ranked based on their information scent value. Images assessed via cues are ranked first in the list of interesting items if they have highest information scent value, and the rest of the images are ranked in the decreasing order of their information scent score. We computed the information scent score based on~\cite{jaiswal2019effects,loumakis2011image}. 

The user study involves finding recommendations on categories such as \say{noodle}, \say{spaghetti bolognese}, \say{painting}, \say{sketches} and \say{landscape} etc., all described in the following section. Users' have been given two options: first is to follow images via cue and the second way is to pursue images through user preferences. All of the participants have been given ten iterations to find q recommendation for each given image category. We ordered the list of \say{interesting} and \say{uninteresting} items by averaging the results from ten iterations based on information scent score.

\section{Experimental Evaluation}
\subsection{Data Collection}
To evaluate the information scent artifacts in our proposed recommendation system we collected a real image dataset from Pinterest.com, a popular visual discovery sharing platform. We collected over 1,116 images belonging to two categories of foods which includes \emph{Spaghetti Bolognese} and \emph{Zoodles}. We split the image data into 67\% train and 33\% test data. Thus, we label those images which users are interested in as \say{1} and uninterested as \say{0}, making it a binary prediction problem. The associated information labels such as title and description of the images may indicate a very complex concept, where we use Naive Bayes analysis to count the frequency of keywords (after data cleaning process). This implementation process applied for the collected Pinterest images.

\begin{figure}[h!]
  \includegraphics[width=\linewidth]{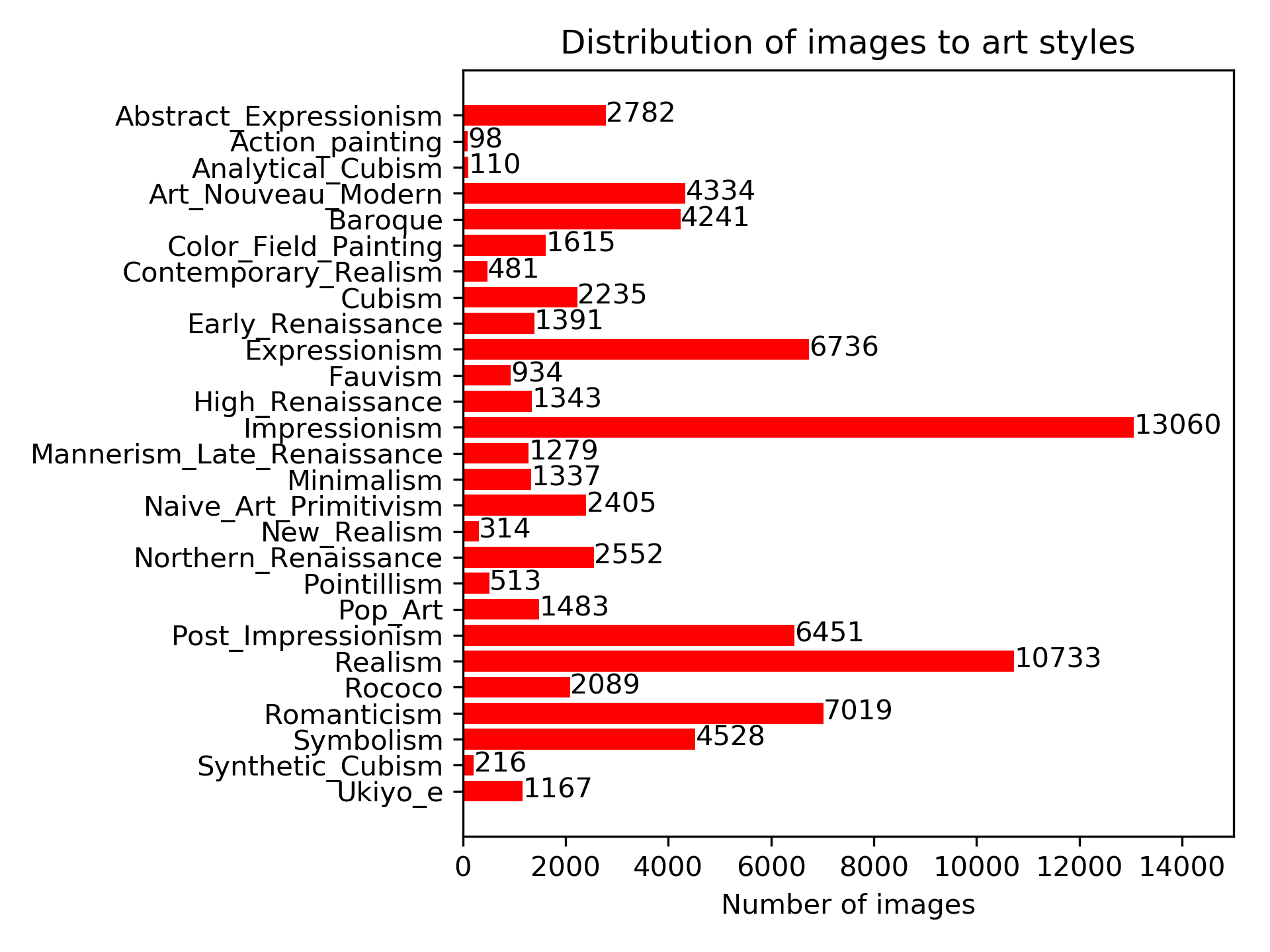}
  \caption{Distribution of images in WikiArt dataset}
  \label{fig:fig_2}
\end{figure}

To test the feasibility of the proposed information scent strategy, we adopt a large-scale collection of art data to provide more human expertise for characterising the images and their varied features during recommendation. We evaluated our image recommendation system on art work which uses the \emph{WikiArt dataset}\footnote{https://github.com/cs-chan/ArtGAN/tree/master/WikiArt Dataset} which contains over 80,000 images of art work labeled across 27 varied art styles collected from WikiArt.org shown in Figure~\ref{fig:fig_2}. 
 \begin{footnotesize}
\begin{table*}[!htb]
    \caption{Classification Report}
    \centering
    \begin{tabular}{|c|c|c|c|c|c|c|c|c|c|}\cline{3-10}
    \hline
    \multirow{4}{*}{\textbf{\diagbox{Class}{Model}}} &
    \multicolumn{6}{|c}{\textbf{Pinterest Collection}} & \multicolumn{3}{|c|}{\textbf{WikiArt Dataset}} \\ \cline{2-10}
     & \multicolumn{3}{|c}{\textbf{GS-SVM}} & %
        \multicolumn{3}{|c}{\textbf{GS-Random Forest}} & \multicolumn{3}{|c|}{\textbf{XGBoost}} \\ \cline{2-10}
    & \multicolumn{9}{c|}{Scores}\\
    \cline{2-10}
    & Precision & Recall & F1 & Precision & Recall & F1 & Precision & Recall & F1  \\
    \hline
    uninterested~(0) & 0.77 & 0.85 & 0.81 & 0.80 & 0.89 & 0.84 & 0.81 & 0.62 & 0.70 \\
    \hline
    interested~(1) & 0.81 & 0.70 & 0.75 & 0.85 & 0.74 & 0.79 & 0.47 & 0.53 & 0.50 \\
    \hline
    \end{tabular}
    \label{tab:tab_1}
\end{table*}
  \end{footnotesize}
  
We use varied image features such as color, content including shape~(width and height) and image quality~(aspect ratio) to characterise. Since our early image recommendation system prototype lacks handling such a large collection of images, we took a small subset of images from the WikiArt dataset~\cite{saleh2015large} and took 1k images as train set and approximately 500 images as test set from over ten different categories, which are abstract\_painting, cityscape, genre\_painting, illustration, landscape, nude\_painting, portrait, religious\_painting, sketch\_and\_study, and still\_life. These ten image subclasses are selected from all 27 different art image categories. Our ten subclasses are chosen to provide conciseness to the user. We use a pre-trained ResNet34 model~\cite{he2016deep} to train our content classifier~(trained on each of the above mentioned image features) which classifies over 1,500 images. We train a tree-based model - XGBoost as a classifier to find the prediction labels i.e., \say{interested} and \say{uninterested}. We train the model at a learning rate of 1e-3 for over 30 epochs. The target labels are the ten subclasses of images where we separate \say{interested} and \say{uninterested} images based on the classification accuracy. We found only four subclasses of images~(illustration, nude\_painting, still\_life, and abstract\_painting) as \say{interested} which results in high classification accuracy~(66\%, 65\%, 57\% and 54\%). As mentioned earlier, the users' interested images are in the given subclasses with corresponding accuracy in order. On the contrary, the users' uninterested images belong to the rest of the six subclasses with classification accuracy in order as ['landscape': 14\%, 'cityscape': 28\%, 'religious\_painting': 32\%, 'sketch\_and\_study': 35\%, 'genre\_painting': 36\%, and 'portrait': 37\%]. The prediction labels support the information scent effects considered during estimating the classification accuracy in a way that art work images takes considerable amount of effort in selecting the recommended images.
\vspace*{-2mm}
\subsection{Results}
 
This section details the evaluation result of implicit behavioural signals in form of information scent scores consistent to the binary prediction problem, where \say{0} and \say{1} signifies weak information scent and strong information scent, respectively. We performed fine-tuning using GridSearch~(GS) on SVM~(Scalable Vector Machine) and Random Forest classifiers with classification scores reported in Table~\ref{tab:tab_1}.
\begin{figure}[h!]
  \includegraphics[width=0.5\linewidth]{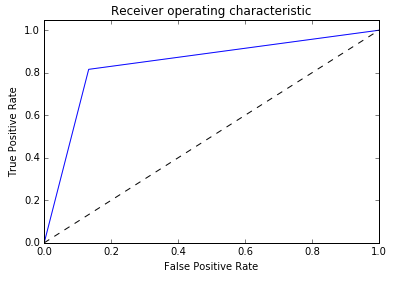}
  \caption{Area under the ROC curve}
  \label{fig:fig_3}
\end{figure}
The evaluation result on WikiArt dataset is reported in the last column of Table~\ref{tab:tab_1}.

The area under the ROC curve~(AUC) measures the rank of predictions to the rank of correct class labels which is reported in Figure~\ref{fig:fig_3}.

In IFT perspective, an image having either "\say{Zoodles} or \say{Bolognese}~(as in Figure~\ref{fig:pgui}) emits strong information scent as a user perceives recommended items which are attended via visual cues. This means that such an image, conferred as information patches (rectangles drawn on images in Figure~\ref{fig:pgui}), delineates significantly a large degree of perception by the user while likely to consume maximum information~(information diet) and having lower information access costs, for instance in terms of the time spent on search.

Our recommendation system is evaluated on the Pinterest collection and the WikiArt dataset. The recommended list of images to users are facilitated via cues~(visual, textual, etc.) and ranked via their information scent score after the user study. We further this work in a way to study if our recommendation system can be strengthened by the additional implicit signals. To check the effectiveness of our proposed approach based on IFT, we perform classification on the image collections used during recommendation to see if the addition of implicit factors improvise the users' interesting and uninteresting items. We report the classification result in Table~\ref{tab:tab_1}. We found that the F1-score of the \say{uninterested} collection of images is significantly higher than the \say{interested} list of items which is due to the diminishing return of information scent score~(in line with~\cite{azzopardi2017building}). It means that the user selected item in first glance put less effort~(selection cost) and those followed several iteration on user preferences to select an image results in weak information scent~(and ranked lower in the list of interesting items). The F1-score on WikiArt dataset is worse due to the diversity of mixed colors and unspecified objects within image patches. We intend to further this work on characterising features at patch-level so that the significance of image during recommendation can be improvised.

\vspace*{-4mm}
\section{Conclusion and Future Work}
This study has utilised a behavioural information retrieval theoretical framework known as Information Foraging Theory in a content-based image recommendation scenario to find implicit features such as the presentation context of a content~(such as an image). Our study found that implicit features such as visual cues~(or bookmarks) in Pinterest~(Pins) can be incorporated with images collectively, with the latter used to recommend images. Also, the implicit features drawn from a subset of the WikiArt image collection are shapes and colors in particular. We found these features based on the varied image size in a diverse image collection of artwork. The larger the size of an image in the recommended lists which exhibits high information scent as artwork images contain various scriptures with fine-grained features that conjointly enhance the user perception, which is likely to be recommended to and consumed by the user. These features help users avoid irrelevant consumption of extra information diet~(by memorising either items or buttons/tags). This leads to minimise the problem of information overload as well as shape their cognitive beliefs while finding interesting items. In future, we intend to apply our proposed method in explainable recommendation for cases found in the evaluation of WikiArt dataset by interpolating the random size of images as information patches.

\bibliographystyle{ACM-Reference-Format}
\bibliography{acmart}
\end{document}